\documentclass[12pt,tightenlines,eqsecnum,floats,shownopacs,nofootinbib,amsmath,amssymb,aps,prd]{revtex4}

\usepackage{color}
\usepackage{graphicx}
\usepackage{amsmath,amssymb}

\usepackage{setspace}
\usepackage{enumerate} 
\usepackage{colordvi} 
\usepackage{subfigure}

\def\be{\nopagebreak[3]\begin{equation}}
\def\ee{\end{equation}}
\def\ba{\nopagebreak[3]\begin{eqnarray}}
\def\ea{\end{eqnarray}}

\def\SU(2){\rm{SU(2)}}
\def\su(2){\rm {su(2)}}

\def\cm{\rm cm}

\begin{document}

\title{The Last 50 Years of General Relativity \& Gravitation:\\ From GR3 to GR20 Warsaw Conferences}
\author{Abhay Ashtekar}
\affiliation{Institute for Gravitation and the Cosmos \&
     Physics Department, Penn State, University Park, PA 16802,
    USA}

\begin{abstract}

This article has a dual purpose: i) to provide a flavor of the
scientific highlights of the landmark conference, GR3, held in July
1962 at Jablonna, near Warsaw; and, ii) to present a bird's eye view
of the tremendous advances that have occurred over the half century
that separates GR3 and GR20, which was again held in Warsaw in July
2013.

\end{abstract}
\maketitle

\section{Introduction}
\label{s1}

The 1962 GR3 conference in Warsaw/Jablona was the last GRG event
before the semi-centennial of Einstein's discovery of general
relativity and this conference, GR20, will be the last one before
the centennial. Therefore, the organizers of GR 20 thought it would
be appropriate to open this conference with a reminder of GR3 and a
brief assessment of the evolution of our field since then. Marek
Demianski provided a vivid portrait of GR3 itself. I will discuss
some scientific highlights of GR3 and contrast what we know now with
what we knew then.

GR3 was a scientific milestone in that, thanks to participants like
Peter Bergmann, Hermann Bondi, Subrahmanyan Chandrasekhar, Bryce
DeWitt, Paul Dirac, J\"urgen Ehlers, Richard Feynman, Vladimir Fock,
Vitaly Ginzburg, Leopold Infeld, Andr\'e Lichnerowicz, Achilles
Papapetrou, Nathan Rosen, Dennis Sciama, John Synge, Joseph Weber
and John Wheeler, it sparked new directions of research in
mathematical general relativity, gravitational waves, quantum
gravity and relativistic astrophysics. At GR20, it is hard to
compete with this illustrious gallery of names. But we are doing
better in two respects. First, our field as a whole has evolved
tremendously, becoming prominent in a variety of disciplines;
physics, astronomy, mathematics, computer science and even some
technologies.  Second, GR3 had no plenary talk by a female scientist
while 20\% of our plenary speakers are women, a significantly larger
fraction than the membership of the International Society on General
relativity and Gravitation. I hope this striking success will make
our field even more attractive to all under-represented groups.

The proceedings of GR3 \cite{gr3} are truly outstanding because they
contain not only what was presented in the main talks and seminars,
but also the tape-recorded exchanges that took place between the
participants after these talks and during special discussion
sessions. When I was a graduate student at Chicago, I borrowed them
from the library and kept them for two full quarters. Although the
proceedings were already a decade old, I learned a lot especially
from the stimulating questions, answers and comments during
discussions.

\begin{figure}[tb]
  \begin{center}
  \subfigure[]
      {\includegraphics[width=2.7in]{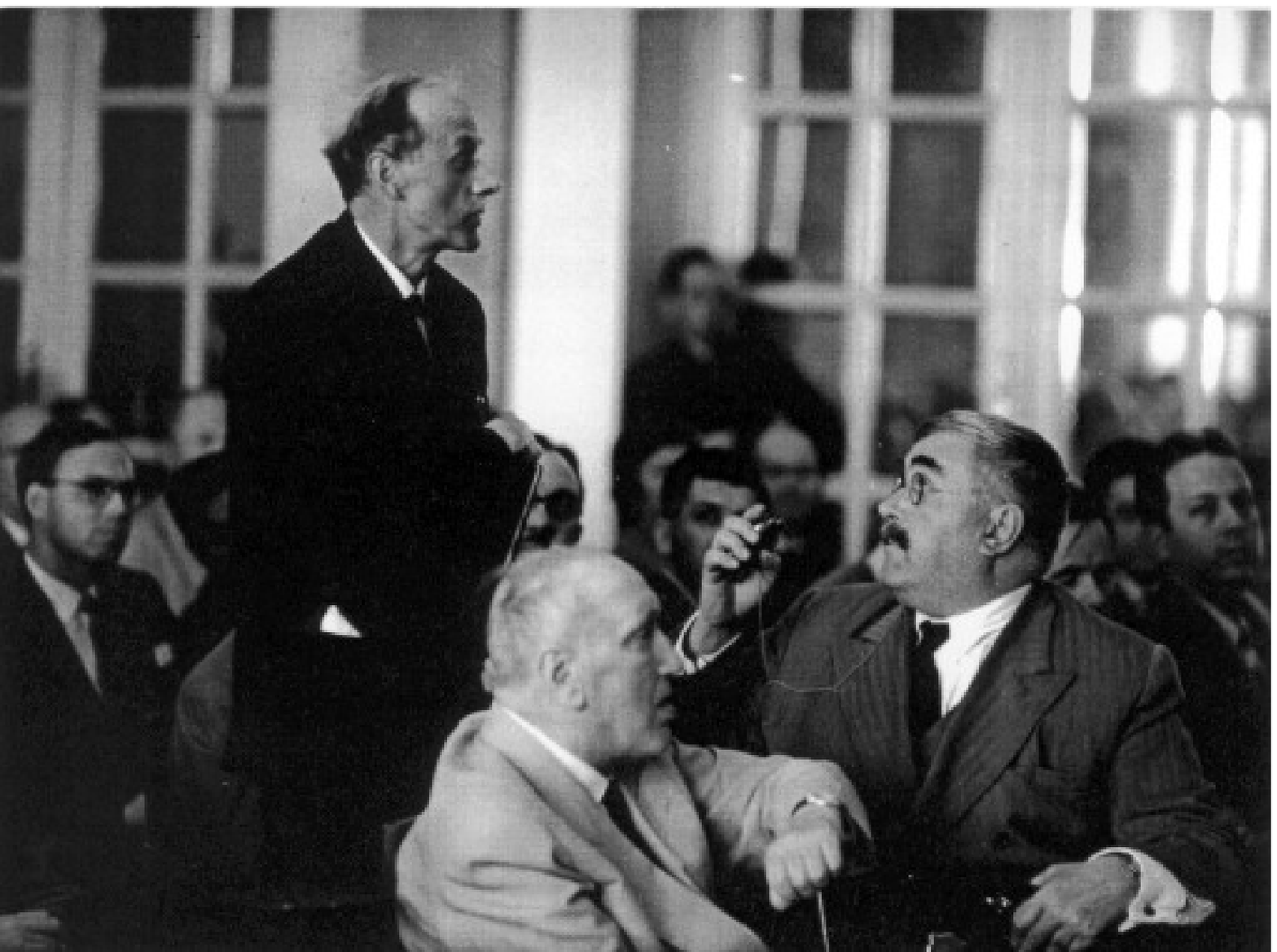}}
      \hskip0.6cm
    \subfigure[]
      {\includegraphics[width=2.5in]{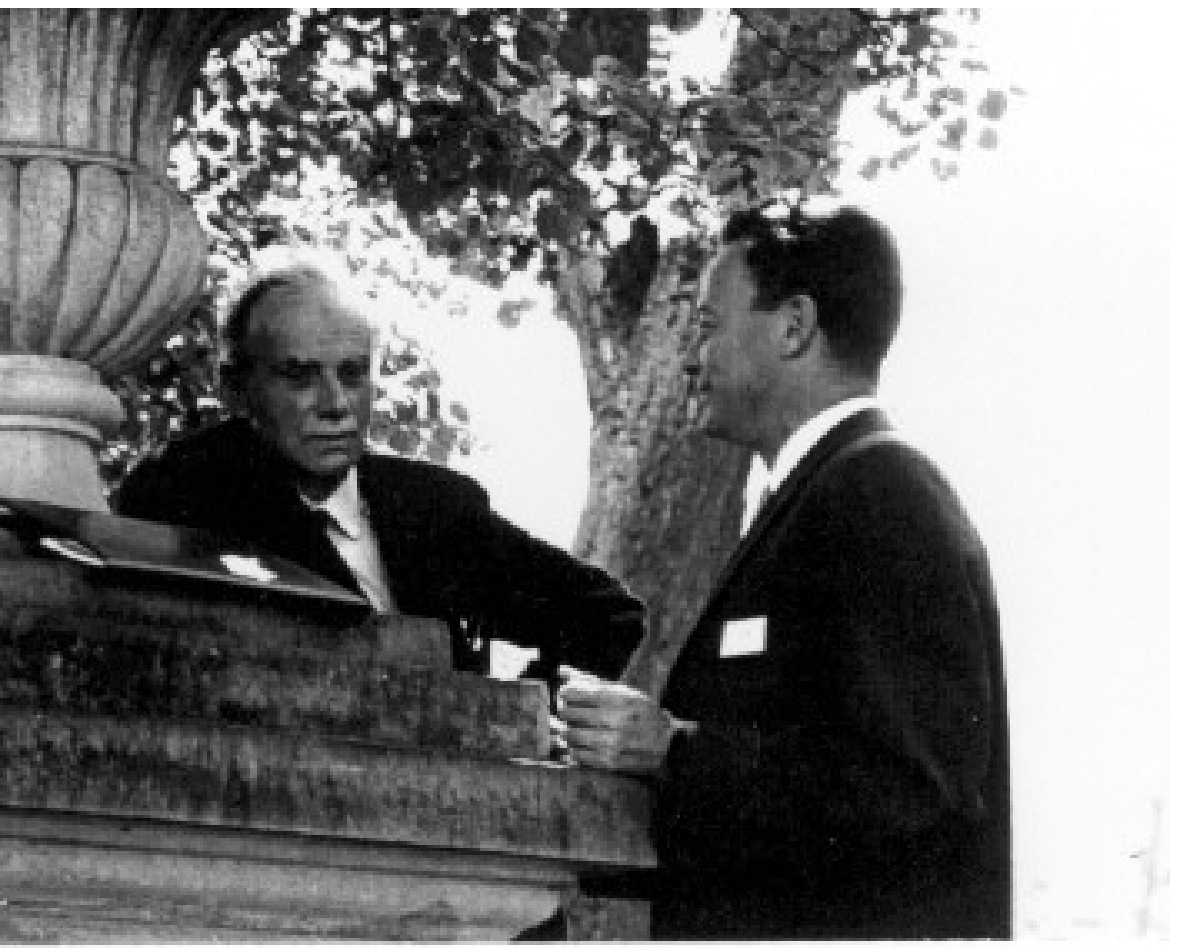}}
    \caption{\hskip0.2cm(a) Dirac, Fock and Infeld, \hskip0.1cm and,
    \hskip0.3cm (b) Dirac and Feynman, \hskip0.1cm at GR3.\\
    \centerline{Credits: Photographs by Marek Holzman (1962).}}
  \label{fig}
 \end{center}
\end{figure}

Let me provide a few examples of these exchanges to illustrate their
general flavor, especially because the younger participants of GR20
would be surprised that so many of the issues that were discussed
have continued to be important ---and in some cases even central---
to our field during the intervening 50 years. On the mathematical
side, Rainer Sachs discussed the characteristic initial value
problem in general relativity, and this was the first conference
proceedings in which Penrose diagrams appeared. Interestingly, Weber
asked why one uses asymptotically Minkowskian boundary conditions in
the study of gravitational waves, rather than asymptotically
Friedmann type. Bergmann and Bondi answered that this is the best
they had been able to do: ``[we] regret that we haven't got to the
point of doing the Friedmann universe." On the experimental side,
Ginzburg and Leonard Schiff proposed the gyroscope experiments to
test the effects of `dragging of inertial frames'. Weber spoke of
the $1000$ kg aluminium cylindrical antenna that had just started
taking data at the University of Maryland, with a sensitivity of
$10^{-14}\, \cm$ in displacements. Feynman asked for the frequency
of gravitational waves that the antenna could detect (and Weber
answered: ``1600 Hz"). Ginzburg was interested in knowing if the
antenna could detect gravitational waves emitted by a binary star
system (and Weber answered: ``For known double stars, No"). On the
quantum front, this was the first conference proceedings  in which
Feynman diagrams with gravitons appeared; Stanley Mandelstam argued
that \emph{``Quantization [of gravity] in flat space can only be
regarded as provisional solution ... one would like to formulate the
equations of the theory exactly, even though approximations have to
be made in their solutions";} and DeWitt spoke of `quantization of
geometry'.

I will divide my remarks into four parts in which the scientific
program of GRG conferences has been traditionally divided for
several decades: General relativity proper, Applications to
cosmology and relativistic astrophysics, Experimental gravity and
Quantum aspects of gravitation.

\section{General relativity, proper}
\label{s2}

As one might expect, this area dominated GR3. Four of the main talks
were devoted to asymptotics and conservation laws and three to
gravitational waves.

On asymptotics, Richard Arnowitt, Stanely Deser and Charles Misner
spoke of new developments in the ADM framework which covered diverse
issues such as gravitational waves, Newtonian limit, singling out
Minkowski space-time, the gravitational vacuum, and transformation
properties of the ADM energy-momentum. The framework they were then
developing has been a driving force behind a variety of developments
over subsequent decades, including numerical relativity, canonical
quantization and positive energy theorems.

The subject of gravitational waves turned out to be a focal point:
this was the first conference in which the reality of gravitational
waves in full general relativity was firmly established. Since the
younger participants at GR20 grew up in the LIGO-Virgo era, it may
come as a surprise to them that the scientific community was
ambivalent about gravitational waves for half a century after
general relativity was discovered. Let me therefore go back in the
years and briefly explain the situation. Einstein did analyze
gravitational waves in the weak field approximation around Minkowski
space just a year after his discovery of general relativity.
However, there was considerable confusion on the subject largely
because people could not separate coordinate effects from true
physics. Eddington for example did not believe in gravitational
waves in full general relativity and famously said that ``they
traveled with the speed of thought". Even Einstein contributed to
the confusion because he first misinterpreted his results with Rosen
on what is now known as the Einstein-Rosen cylindrical waves!%
\footnote{For a step by step account of this fascinating episode,
see Kennefick's 2005 talk \cite{er}.}
Indeed, he wrote to Max Born in mid-1936:
\begin{quote}
\emph{``Together with a young collaborator I arrived at the
interesting result that gravitational waves do not exist, though
they had been assumed to be a certainty to the first approximation.
This shows that non-linear gravitational wave field equations tell
us more or, rather, limit us more than we had believed up to now."}
\end{quote}
Such misunderstandings were set to rest in the early 1960's by a
group of researchers, many of whom were associated with Bondi's
group in London. Bondi considered space-times describing isolated
systems that represent sources of gravitational waves and recognized
that the physical properties of gravitational waves could be teased
out of these geometries by moving away from the sources in (future
pointing) \emph{null directions}. By encoding the energy carried
away by gravitational waves in an invariant field ---now called the
Bondi news--- he dispelled the confusion between coordinate effects
and true physics. As he often emphasized later, `gravitational waves
are real; one can boil water with them!'

But even after these announcements, some confusion still persisted
at GR3. In particular Bergmann raised the question as to
\emph{``whether an n body system held together by purely
gravitational forces will radiate or not."} Feynman provided an
affirmative answer using perturbative, quantum scattering theory.
But not every one was convinced, again because of the concern that
Feynman's weak field calculations ignored the infinite dimensional
diffeomorphism freedom of the full theory. Also, in his final
summary of the conference as a whole, Bergmann emphasized that ``the
rate of energy (or linear momentum) radiation will depend on the
choice of coordinate system,
and change under supertranslations in an involved manner". 
If this were the case then the energy carried away by gravitational
waves would not be unambiguous.

Fortunately, thanks to further work by Ted Newman, Roger Penrose,
Sachs and others this concern turned out to be unfounded. A clear
theory of gravitational waves in \emph{full general relativity}
emerged in the asymptotically Minkowskian context. Together with
subsequent work on approximation methods, such as the post-Newtonian
scheme and increasingly reliable analysis of equations of motion,
this theory provides the foundation for much of the current work in
numerical relativity as well as data analysis in the gravitational
wave science.\\

Since GR3, the area of mathematical general relativity has made
tremendous advances in many other directions as well. I can only
include a few highlights to illustrate the diversity of these
successes since an inclusive summary of even the major areas will
exceed the allocated space.

First, at the time of GR3, there was still controversy as to whether
singularities in the known exact solutions were artifacts of high
symmetries. The Khalatnikov-Lifshitz school had a program to show
that the general solution of Einstein's equations would be
singularity free generically. What cleared up the situation was the
introduction and astute use of global techniques by Penrose. The
first singularity theorems by him, Stephen Hawking, Robert Geroch
and others established that singularities are not restricted to
symmetric solutions but arise generically in physically interesting
situations. However, these theorems assumed that matter satisfies
certain energy conditions, which are violated by the scalar fields
used in inflation. So there was a hope that they could be evaded and
in the inflationary scenario there would be no big bang in GR. But
subsequently Arvind Borde, Alan Guth and Alex Vilenkin established
more general results to show that this is not the case: In GR the
big bang singularity is inevitable. The behavior of space-time
geometry as one approaches such space-like singularities has drawn
considerable attention. By now there is impressive analytical and
numerical evidence for the Belinskii-Khalatnikov-Lifshitz conjecture
that `time derivatives dominate over spatial ones' in this limit.

A second direction that is even more significant in terms of global
problems was opened up by the positive energy theorems proved by
Richard Schoen, Shing-Tung Yau, Edward Witten and others in the late
seventies and early eighties. For physical fields in Minkowski space
we know that the total energy momentum is causal and future
directed. However, the issue becomes less clear-cut once the
gravitational field in included because the gravitational potential
energy is negative. Could the negative potential energy not
overwhelm the positive matter contributions in the strong field and
highly non-linear regime of general relativity? Indeed, such
concerns were expressed already at GR3. The positive energy theorems
cleared up this issue. They established that, so long as the matter
sources have a future directed, causal 4-momentum density, the total
(Arnowitt-Deser-Misner) 4-momentum, as well as the 4-momentum at any
retarded time instant (the Bondi 4-momentum) are also future
directed and time-like. Furthermore, these theorems provided a new
invariant for asymptotically flat Riemannian manifolds that is of
interest in own right to differential geometry. Through this work,
Schoen and Yau introduced non-linear geometric analysis
---the area that forms the interface of geometry and partial
differential equations--- to the general relativity community and
general relativity to mathematicians. A whole new generation of
mathematicians was thus attracted to global problems in general
relativity. Using geometric analysis, non-linear stability of
Minkowski space and deSitter could be established by Demitrios
Christodoulou, Sergio Klainnerman, Helmut Friedrich and others.
Finally, as we heard at GR20, the astute combination of geometric
analysis and numerical techniques has now led to the surprising
result that anti-deSitter space-time is non-linearly
\emph{un}stable.

The third area was not even discussed at GR3 but has seen a true
explosion of activity since then:  black hole physics. Just after
GR3, Roy Kerr generalized the Schwarzschild solution and its physics
was understood by a systematic analysis of its global structure by
Brandon Carter and Penrose. Hawking's introduction of event horizons
enabled a precise formulation of the notion of general black holes.
A major surprise came through the black hole uniqueness theorems of
Werner Israel, Hawking, David Robinson, Pawel Mazur, G. L. Bunting
and others. While there is a very large variety of stationary stars,
stationary black holes of astrophysical interest are tightly
restricted. Furthermore, space-times describing them are
\emph{explicitly known}, provided just by the exact solutions of the
Kerr family! This result opened the rich field of perturbations of
Kerr space-times. Finally, there was the surprising discovery by
James Bardeen, Carter and Hawking that there is a close analogy
between the laws of black hole mechanics and the laws governing
ordinary thermodynamics, and the simultaneous analysis of this
interplay through thought experiments by Jacob Bekenstein. These
results brought out the deep and awe inspiring unity between general
relativity, statistical mechanics and quantum physics, and helped
attract high energy physicists to gravitation. The conceptual issues
that opened up continue to lie at the forefront of fundamental
physics.

The last area I want to discuss did not even exist at the time of
GR3: Numerical relativity. Introduced by DeWitt and Larry Smarr in
the early seventies, it has truly blossomed during the last decade.
It has provided brand new insights into the full non-linear regime
of general relativity. On the conceptual side, we learned that there
is unforeseen critical behavior associated with the gravitational
collapse, first discovered by Matthew Choptuik. It came with
structures that are standard in statistical mechanics but completely
new to gravitational physics: scaling behavior, critical indices and
universality. On the mathematical side, through the work of Bernd
Bruegmann, Frans Pretorius and others, numerical relativity has
finally led to a solution of the famous 2-body problem of general
relativity. On the astrophysical side, numerical techniques taught
us that the gravitational collapse and inspirals are much tamer than
what had been assumed till then: there are no spin flips as black
holes merge, no novel signatures of non-linear dynamics of the last
plunge. But not everything was foreseen by approximation methods.
For example numerical simulations revealed that, because of the
gravitational wave emission, the final black hole resulting from a
coalescence \emph{can} receive a very substantial kick. Numerical
results also pressed upon us to better understand dynamical black
holes. This led to a new bridge between analytic and numerical
methods through the introduction of quasi-local horizons,
definitions of mass, angular momentum and multipole moments
associated with them and balance laws for these quantities. While
the original motivation for this work was to provide invariant tools
to extract physics from numerical simulations in the fully
non-linear and dynamical regimes, it has led to interesting results
in other areas. In geometric analysis, existence, uniqueness and
dynamics of marginally trapped surface were investigated, and new
inequalities on angular momentum, reported at GR20, were discovered.
Quasi-local horizons have also provided physically interesting
generalizations of black hole thermodynamics by dropping the
assumption of global stationarity, and precise mathematical tools
that are needed to track the dynamics of black hole evaporation.\\

Perhaps the most striking aspect of all these developments is that
they have brought out ---and continue to bring out--- the deep
underlying currents that connect areas of physics, astronomy and
mathematics that had been seen as disparate for decades. There was
no inkling of any of these multifaceted interconnections at the time
of GR3.

\section{General Relativity and the Universe}
\label{s3}

\subsection{Cosmology}
\label{s3.1}

It is very difficult for younger researchers to fully appreciate
\emph{how much} this field has progressed. Because the observational
data were so limited, at the time of GR3 the subject was dominated
by philosophical considerations and a majority of European
--especially British--- physicists preferred the steady state
scenario over the big bang paradigm. In fact the term big bang was
meant to be pejorative. It was coined by Fred Hoyle, the strongest
proponent of the steady state model, to poke fun at the idea of a
finite beginning in a big explosion! At GR3 there was not a single
main talk, and only two seminars, on cosmology proper. One, by
Bondi, was on steady state cosmology while the second by A. L.
Zelmanov was on anisotropy and inhomogeneity.

During the last several GR conferences, by contrast, theoretical and
observational cosmology have been major focal points. This
transformation has come about primarily because of the spectacular
advances on the observational front, and also because of the influx
of ideas from the theoretical high energy physics community. Now
that the theory has been repeatedly confronted with observations via
completely independent initiatives, the idea that the universe is in
a steady state has been abandoned.

Progress on observational front has been simply spectacular. COBE,
the Cosmic Background Explorer was launched in 1989; WMAP, the
Wilkinson Microwave Anisotropic Probe in 2001; and Planck in 2009.
Most of us were amazed by the accuracy of the COBE measurements when
the results were first announced. It has been surpassed by leaps and
bounds. WMAP reduced the allowed volume of cosmological parameter
space by a factor in excess of $30,000$ by, in particular, improving
the angular resolution $33$ times with respect to COBE. Planck made
a similar jump with respect to WMAP. While WMAP had 5 frequency
bands between $23$GHz to $94$GHz, Planck has nine, between $30$GHz
to $857$GHz; Planck has probed much smaller angular scales where the
error bars are even smaller. Therefore, in the Planck analysis, the
best fit to the six parameter model is obtained using data at these
smaller angular scales. Members of the Planck team often point out
that, in terms of overall precision, what Planck could do in a year
and a half, WMAP would have needed a thousand years of observations
and COBE would have needed a million years of operation.

In addition we have ground based facilities, such as ACT, the
Atacama Cosmology Telescope in Chile, and SPT, the South Pole
Telescope. In addition to these CMB measurements, Type 1a supernovae
(which serve as standard candles) and the Baryonic Acoustic
Oscillations (which provide the standard ruler in cosmology) have
enriched our understanding of the expansion history of the universe
and structure formation. It is a striking fact that all these
largely independent measurements provide a consistent picture of the
early universe and its evolution, compatible with what has now
become the six parameter, standard concordance model in cosmology.

It is interesting to contrast what we know today with what the
community knew during the decade of GR3. In the 1960s and 70s, we
had no clue as to whether the total energy density in the unverse is
close to the critical density, or much less, or much more. Many of
the leading physicists, astronomers and cosmologists preferred a
closed universe ---i.e. critical density greater than $1$--- on
aesthetic grounds, e.g., because it `neatly avoids the issue of
creation'. WMAP taught us that the universe is spatially flat to
within $0.6\%$ accuracy! In 1964 the Hubble parameter was estimated
to be $125 {\rm km}\,\, {\rm s}^{-1} {\rm Mpc}^{-1}$. In the 1970s
there was a lively debate between de G\'erard Vaucouleurs (who
argued for larger value) and Allan Sandage and G. A. Tammann (who
argued for smaller value) as to whether it was $100 {\rm km}\,\,
{\rm s}^{-1} {\rm Mpc}^{-1}$ or $50 {\rm km}\,\, {\rm s}^{-1} {\rm
Mpc}^{-1}$. The first 15.5 months' data from the Planck satellite
has revealed \cite{planck} that $H_0 = (67 \pm 1.2) {\rm km} {\rm
s}^{-1} {\rm Mpc}^{-1}$; the $100\%$ uncertainty is reduced to less
than $1.8\%$! The Planck data also tells us that the Cosmic
Microwave Background (CMB) temperature today is $(2.7255\pm
0.0006)^\circ {\rm K}$, i.e., with a $0.02\%$ accuracy. We are
confident, to $10\sigma$ level, that the power spectrum is
\emph{not} exactly scale invariant. The spectral index $n_s$ for
scalar (or density) modes is $0.9608 \pm 0.0054$ (exact scale
invariance would mean $n_s =1$). At GR3, almost everyone thought
that `normal matter' that constitutes the stars and galaxies,
photons and neutrinos accounted for the entire matter content of the
universe. Today, we know that it contributes negligibly and the
primary drivers of the evolution of the large scale structure of the
universe in the $\Lambda$CDM model are dark matter and the
cosmological constant. The Planck data tells us that the normal
matter that makes up stars and galaxies contributes just 4.9\% of
the mass/energy density of the Universe. Dark matter, which has thus
far only been detected indirectly by its gravitational influence,
makes up 26.8\%, and the rest is `dark energy' best modeled by a
positive cosmological constant. The `standard' paradigm could hardly
have shifted more dramatically!

Progress on the theoretical side has also been impressive. First,
one can use the $1$ part in $10^{-5}$ homogeneities observed in the
CMB to construct the initial data for perturbations on a Friedmann
Lemaitre Robertson Walker (FLRW) background, and evolve them using a
mixture of analytical and numerical methods. These simulations use
general relativity and well established classical physics,
astrophysics and gastrophysics. The result in very good agreement
with the large scale structure that has been observed through
astronomical surveys such as the Sloan Digital Sky Surveys. This is
truly impressive because the evolution starts when the universe was
about $370,000$ years young and end today, over $13$ billion years
later. In human terms it is like taking a snapshot of a baby when
she is a day old and have a reliable biological and biochemical
theory to accurately predict what the person would look like when
she is a hundred years old!

These computations trace back the origin of the large scale
structure to the minute inhomogeneities observed in the CMB. But one
can be an order of magnitude more ambitious and ask for the origin
of the \emph{CMB fluctuations themselves}. The surprise here is that
there do exist candidate theories ---such as the simplest single
field inflation with a quadratic potential--- that rise up to this
challenge.%
\footnote{As Penrose and others have emphasized, contrary to what is
often claimed, inflation does not `solve' the flatness and horizon
problems. For example, given any inflationary potential, one can
just give initial data at a suitably late time with significant
spatial curvature and simply evolve them back in time to obtain
initial conditions at the onset of inflation. When these are evolved
forward, obviously one would find that there is significant spatial
curvature at late times! The power of the paradigm lies, rather, in
its success in accounting for the CMB anisotropies.  (It also solves
the `monopole problem' which, however, is no longer considered as
significant by cosmologists.)}
The paradigm does involve some assumptions whose validity is still
far from being obvious. But the power of the argument is that: i)
these assumptions are few and can be stated precisely; and, ii) if
the quantum fields representing perturbations were in one of a
natural class of `vacua' just before the onset of the slow roll
inflation, then the initial vacuum fluctuations are naturally
magnified during inflation to produce precisely the power spectrum
we observe in the CMB. Thus, the issue of the origin of the observed
large scale structure of the universe is now pushed back all the way
to the onset of inflation, i.e., just a million Planck seconds after
the big bang of general relativity. Furthermore, it is very striking
that the initial perturbations are now reduced just to the
\emph{vacuum fluctuations} of quantum fields. In this scenario, at
the `beginning' the universe was as homogeneous and isotropic as it
is possible, subject to the Heisenberg uncertainty relations that
\emph{must} be obeyed by the quantum fields representing
perturbations. But this paradigm is also incomplete. In particular
one of the assumptions is that one can trust general relativity and
quantum field theory on FLRW space-times all the way to the big bang
singularity. At GR20 we also heard about recent advances in loop
quantum gravity where the goal is to obtain self consistent Planck
scale completions of the leading paradigms of the very early
universe.\\

To summarize, since GR3, there have been \emph{very} significant
advances in cosmology especially on the observational front, but
also on the theory side, that have dramatically changed our view of
the matter content, dynamics and the large scale structure of the
universe. What is even more pleasing is that the pace of progress in
this area is likely to remain in a high gear for the foreseeable
future.

\subsection{Relativistic Astrophysics}
\label{s3.2}

GR 3 did not have a single talk in this area because the field did
not even exist. Soon thereafter, there was an explosion of interest,
again because of an observational breakthrough, the discovery of
quasars, and the field came into official existence with the first
Texas Symposium on Relativistic Astrophysics, held at Dallas in
1963. GR3 was the first general relativity conference that
Chandrasekhar attended and from conversations with him I know that
his decision to devote the next quarter of a century to general
relativity was strongly influenced by this conference. This is very
fortunate for the development of this area because Chandra firmly
believed that the natural home of general relativity is in
astronomy. Over the next 2-3 decades, his group at Chicago and Kip
Thorne's group at Caltech worked tirelessly to develop this
direction.

By now, there are many applications of general relativity to
astrophysics. The possibility of gravitational lenses due to general
relativistic effects was already discovered by Einstein. But he did
not think that they would ever be observed. Not only have they been
seen but they have now become a standard and \emph{powerful tool}
with a wide range of applications, from the search of exoplanets in
astronomy, to the analysis of CMB data in cosmology. General
relativity also plays a key role in the study of compact
astrophysical objects. In particular, isolated neutron stars as well
as those in binaries have been cataloged and their formation and
dynamics have been topics of active research in relativistic
astrophysics. Stability analysis has helped us put constraints on
the equation of state of nuclear matter at these high densities.
More generally, relativistic effects are conceptually important in
the stability analysis of all rapidly rotating stars.

The most sweeping change in this field has occurred in the area of
astrophysical black holes. It may come as a surprise to the younger
participants at GR20 that, up until the \emph{late seventies}, black
holes were not taken seriously in the physics and astronomy
communities. A standard view was: `just because a theory allows
certain solutions does not mean than they are actually realized  in
Nature'. This may seem odd, given that the work by Robert
Oppenheimer and Hartland Snyder on black hole formation through
gravitational collapse dates back to 1939. But the
Oppenheimer-Snyder calculation was carried out using exact spherical
symmetry and the collapsing object was modeled by a sphere of dust
with uniform density. This appears to have led to a general
perception that the calculation did not apply to realistic stars.
Indeed, even within the general relativity community a similar view
arose immediately after Israel's discovery that a static black hole
had to be spherically symmetric. At first, this result was
interpreted as implying that black holes would be very rare: The
final result of a gravitational collapse can be a black hole
\emph{only} if the collapsing star is spherical and most stars are
not; they have higher multipoles. It was Penrose who forcefully
argued that the higher multipoles would be radiated away and the
generic stellar collapse of sufficiently massive stars would in fact
lead to black holes. Subsequently, this view was confirmed by
perturbative calculations by Richard Price and others, carried out
between early 1970s to early 1990s. Generalization of those results
to full non-linear general relativity, and the issue of stability of
the Kerr family of black holes singled out by the uniqueness
theorems, lie at the forefront of research in geometric analysis I
referred to in section \ref{s2}.

As more astrophysical phenomena were discovered, by 1990s the tide
turned $180^\circ$ and black holes became ubiquitous in astronomy.
While in the 1974 AAS meeting in Tucson, only Thorne's group was
using black holes to model Cyg X1, it became a \emph{norm} to assume
that smaller black holes of a few solar masses are the engines
behind the spectacular outbursts of energy, e.g., in gamma ray
bursts and supermassive black holes in active galactic nuclei and
quasars. It is now common to suppose that most galaxies have
supermassive black holes at their centers and they are dominant
players in determining the dynamics of their host galaxies. This
relation and the mechanism of formation and growth supermassive
black holes continues to be a frontier topic in astronomy. At GR20
we learned that some of these supermassive black holes were formed
\emph{very} early, requiring us to revise the scenarios of birth and
dynamics of these huge black holes. We also heard about Sgr A*, the
$\sim 4$ million solar mass black hole at the center of our own
galaxy. Interestingly, one has been able to deduce its
characteristics, particularly the mass, from a careful observation
of stars orbiting it, following the proposal John Mitchell made in
1783, in the very first paper on `black holes' \cite{jm}! In the
near future we will have interesting information on the accretion
process into Sgr A* because there is a unique opportunity to observe
a gas cloud of several earth masses as it is devoured by the central
black hole.\\

In summary, the role of general relativity in astrophysics has
magnified very significantly since the Texas symposium. Aspects of
relativistic gravity now play an essential role in understanding
compact astrophysical objects, the most energetic phenomena in the
universe, and the interplay between supsermassive black holes and
large scale dynamics.

\section{Gravitational Experiments}
\label{s4}

At GR3, there were three talks on experimental gravity that I have
already alluded to in section \ref{s1}; one by Ginzburg on tests of
general relativity, one by Schiff on the gyroscope experiment and
the report by Weber on the Maryland gravitational wave antenna.
After reviewing the then status of the three solar system tests of
general relativity,  Ginzburg emphasized: \emph{``There are no
clear-cut experimental results or observations which would negate
G.R.T [general relativity theory]. There is not even the smallest
cloudlet on the horizon ... (of course this refers to macroscopic
physics).. It applies equally well even to cosmology where the
necessity of some generalization or the other would not cause any
particular surprise (the simplest such generalization is the
addition of a $\Lambda$ term ...)".} But he also emphasized that
there were ten other relativistic gravity theories and so it is
important to carry out tests. Ginzburg and Schiff were prescient in
emphasizing that satellites will play an important role in these
tests, particularly that of the Lense-Thirring effect. Apparently,
Einstein had said to Thirring that it is unfortunate that the effect
is so small for the moon and he wished the moon were closer. Schiff
pointed out that we now have many moons (i.e. satellites) that
\emph{are} closer! In his conference summary, Bergmann recalled
Einstein's sentiment that because the foundation of the theory lies
in the equivalence principle, testing it more accurately was  more
important than the solar system tests. He looked forward to the
significant improvement of the E\"otv\"os experiments that were
expected at that time from Robert Dicke's experiment at Princeton.

There was a flurry of activity in the mid-eighties when a reanalysis
of the E\"otv\"os experiment suggested that here may be violation of
the equivalence principle indicating a novel fifth force operating
at the scale of $10^4 \cm$. But the numerous new experiments that
were performed soon ruled out this possibility. By now the
equivalence principle has been tested to one part in $10^{14}$!
There have been impressive tests of Newton's inverse square law down
to $10^{-1} \cm$. The local Lorentz invariance underlying general
relativity has been tested to few parts in $10^{22}$. Over half a
dozen post-Newtonian parameters that distinguish general relativity
from other gravity theories have been measured and agreement with
general relativity has been established in the range of one part in
$10^3$ to one part in $10^{20}$. With respect to the important post
Newtonian parameters $\beta$ and $\gamma$, there has been an
improvement by a factor of 100 over the last 40 years. These tests
have brought out the fact that it is \emph{extremely} difficult to
non-trivially modify or augment general relativity.

The evidence for validity of general relativity from gravitational
phenomena outside the solar system is in many ways even more
impressive. The most celebrated among these observations is the
continued monitoring of the Hulse-Taylor pulsar PSR1913+16 where
general relativity has been confirmed to an accuracy of two parts in
$10^4$. In 2004, a double pulsar PSR J0737-3039 A/B  was discovered.
In addition to being a binary in which pulses from both neutron
stars are received on earth, it is also the most relativistic binary
pulsar observed so far. It has already provided accurate
measurements of masses ---$m_{\rm A} = 1.339\, {\rm M}_{\bigodot}$
and $m_B = 1.250\, {\rm M}_{\bigodot}$ with a $0.2\%$ accuracy--- as
well as six post-Keplerian parameters. They have also enabled the
first quantitative measurement of general relativistic spin
precession. An interesting development is the \emph{Einstein at
Home} project in which systematic searches for pulsars are carried
out using the immense computational power made available by some
40,000 volunteers who have offered the use of some 200,000 personal
computers during periods in which they are idle. This initiative has
led to the discovery of six binary pulsars, which has been credited
to the \emph{``citizen scientists"} who made it possible through
their generosity.

Finally, there is a revolution waiting in the wings that will
ultimately prove to be more important than all these tests:
Gravitational waves. A global network of detectors is being set up.
The LIGO detectors in the US and the Virgo detector in Europe are
undergoing transitions to the `advanced stage' when the sensitivity
would be sufficient for observation of several events a year.%
\footnote{It is interesting to note that in a discussion session at
GR3,  Bergmann prefaced a discussion of a theoretical issue by
saying \emph{``it is perfectly safe to argue the point on
theoretical grounds because Weber is a comfortable number of orders
of magnitude away from deciding this question experimentally; and we
will probably all be dead by the time the decision is in."} There is
every expectation that gravitational waves will finally be seen in
this decade.}
By the end of this decade they will be joined by the Kagra cryogenic
detector in Japan. The LIGO-India observatory will be at a
sufficiently different latitude from the first four to significantly
enhance the source localization and polarization measurement
capability of the global network. Technological advances and the
ingenuity of experimentalists have been mind boggling. Whereas at
GR3, Weber spoke of measuring root mean square displacements of
$10^{-14}\cm$ at ends of an antenna that was $2$ meters long,  LIGO
is capable of measuring displacements of $10^{-16}\cm$ above the
noise floor level, between mirrors that are separated by a distance
of \emph{$4\, {\rm km}$}! The global computational power devoted to
these searches is also impressive. Finally, since gravitational
waves interact \emph{so} weakly with everything, they should bring
to us faithful signatures of the cataclysmic events all the way from
the edge of the observable universe. It is widely expected that this
new window on the universe will have a transformative effect on
astronomy. Indeed even now the upper limits on gravitational waves
set by the LIGO-Vergo collaboration have provided novel insights.
For example, the \emph{absence} of any gravitational wave signal at
the sensitivity that these detectors had already achieved tells us
that there is no mountain on the crab pulsar that is higher
than a meter!\\

While the advances on the experimental front have been significant,
so far they have not been as dramatic as those in other areas I have
discussed. Nonetheless, it \emph{is} quite astonishing that not only
has general relativity withstood such a wide variety of
observational tests but, for almost a century now, it has been
impossible to modify it and come up with interesting, viable
alternatives. Furthermore, the theory predicts that a time changing
quadrupole is accompanied by ripples in space-time curvature. It
would be surprising if the news they will bring from the far corners
of the cosmos does not deepen our understanding of the structure and
dynamics of astrophysical systems in unforeseen ways.

\section{Quantum Aspects}
\label{s5}

At GR3, there were four main talks on quantum aspects of gravity by
DeWitt, Feynman, Madelstam, and Lichnerowicz, and well over half a
dozen seminars were motivated by these issues. In section \ref{s1},
I mentioned the first three talks. In the fourth, Lichnerowicz
discussed his proofs of the existence and uniqueness of various
Green's functions on globally hyperbolic, curved space-times,
notably the commutator Green's function that provides the point of
departure for quantum fields in curved space-times. He covered not
only the simplest case of scalar fields, but also spinor fields and
the linearized gravitational field. It is impressive to see that the
subject was already so advanced. However, he also thought that one
could similarly introduce a canonical notion of positive and
negative frequency decomposition in a general globally hyperbolic
space-time. Work by Parker and others soon proved that this
possibility cannot be realized: in a time dependent space-time,
there is no canonical vacuum and no objective notion of particles.
This understanding subsequently led to the algebraic approach which
we heard about at GR20. By now the theory is mature. From a
conceptual and mathematical viewpoint, it is at the same level as
interacting quantum field theory in flat space-time.

The most significant development in this area is of course Hawking's
discovery that, in the external field approximation, black holes
radiate quantum mechanically and at late times the radiation is well
modeled by a black body. This discovery provided a \emph{physical}
basis to black hole thermodynamics discussed in section \ref{s2}.
Key researchers in the field had expressed the view that the
similarity between the laws of black hole mechanics and
thermodynamics investigated by Bekenstein was just a nice
coincidence. The discovery of the black hole radiance shifted the
entire paradigm, providing a fresh perspective and raising new
questions at the interface of general relativity and quantum
physics. As we saw in GR20, the ensuing issues are still at the
forefront of current research.

The second striking application of quantum field theory in curved
space-times is to the early universe. In this epoch, space-time
appears to be extremely well approximated by a homogeneous isotropic
(FLRW) solution to Einstein's equations, with tiny perturbations.
These are best represented by certain quantum fields propagating on
the FLRW background. As we saw in section \ref{s3.1}, by modeling
perturbations as `vacuum fluctuations' of these quantum fields at
the onset of inflation and then evolving them using quantum field
theory in curved space-times, one can reproduce the inhomogeneities
observed in the CMB. In this sense, cosmology of the early universe
has already provided an observational confirmation of quantum
effects gravity, at the level of linear perturbations.

At GR3, the emphasis was on full quantum gravity, beyond the
external field approximation. On the conceptual side, L\'eon
Rosenfield argued that one could leave the gravitational field
classical while Frederic Balinfante pointed out that it would be
inconsistent to have a fundamental theory in which a classical
metric is coupled to the expectation value of the matter stress
energy tensor. DeWitt and Feynman discussed perturbative quantum
gravity. There was an interesting exchange between the two. DeWitt
expressed the hope that a finite number of counter terms could
suffice and quantum gravity would be perturbatively renormalizable
\emph{if} one used dressed propagators and not bare ones. Feynman
said that he sees that there is a finite number of counter terms at
the one graviton loop level, but he did not see what happens when
there are two or more loops: \emph{``As far as I can see the
gravitational theory is not renormalizable in the usual sense of the
term."} Detailed calculations in the mid-eighties by Goroff and
Sagnotti showed that Feynman's view was correct: Perturbative
general relativity off Minkowski space-time fails to be
renormalizable at the two loop level. But at GR3 there was also
considerable discussion of non-perturbative quantum gravity, in the
canonical framework by Bergmann and others and through path
dependent observables by Madelstam. Also, DeWitt pointed out that
topology change is possible in non-perturbative quantum theory while
David Finkelstein suggested that elementary particles may be
regarded as bound states with non-trivial topologies, the geons.

Interestingly, it is possible to trace back most of the main
directions of contemporary research in quantum gravity to GR3. The
Feynman-DeWitt perturbative approach naturally led to supergravity
and culminated in perturbative string theory. There has been a
resurgence of interest in supergravity over the last decade because
of the discovery that the 4-dimensional, $\mathcal{N}=8$ maximal
supergravity theory is \emph{much} better behaved in perturbation
theory than was expected. This realization has even led to the
conjecture that the theory may be perturbatively \emph{finite}. In
another development, non-renormalizability of perturbative quantum
general relativity led Steven Weinberg to suggest that perhaps
general relativity may be asymptotically safe, i.e., may admit a
\emph{non-Gaussian} fixed point. The asymptotic safety program is
now being actively pursued and impressive evidence in its support
has steadily accumulated over the past decade. The broad framework
of asymptotic safety also underlies the `causal dynamical
triangulation' approach. Finally, Loop quantum gravity can be
regarded as the culmination of both sets of non-perturbative ideas
discussed at GR3: the Dirac-Bergmann-Wheeler canonical quantization
program and Madelstam's approach in which path dependent,
Wilson-lines type observables are at the forefront.

Over the past two decades there have been significant developments
in this area, particularly in string theory and loop quantum
gravity. Even though we are still far from a complete theory in
either approach, both avenues have led to concrete advances by
removing several of the conceptual and mathematical road-blocs the
field faced in 1980s and 1990s. It is quite striking that, in both
approaches, the fundamental building blocks of space-time are very
different from what a simple minded extrapolation from field
theories in Minkowski space-time would suggest. Furthermore these
fundamental excitations appear to be 1-dimensional, polymer-like.
But beyond these basic similarities, the two paths diverge. In
string theory, the original goal was to achieve \emph{unification}
of all forces of nature. While the scope of the theory has become
more diffuse over the years, the unification theme has had a strong
influence on its underlying structures. By contrast, in loop quantum
gravity one focuses on the fact that gravity and space-time geometry
are intimately intertwined. Consequently, \emph{quantum geometry}
effects underlie all the major developments in this approach to
quantum gravity. Over the past decade, the paths have diverged even
more significantly. In string theory, because of the tremendous
success of the ADS/CFT conjecture, the emphasis has been on using
parts of gravity theory that we understand well to explore
properties of strongly coupled systems in \emph{non-gravitational}
areas of physics: properties of the quark gluon plasma, issues in
fluid mechanics, problems in condensed matter physics, particularly
various aspects of superconductivity, \ldots. On the other hand, in
loop quantum gravity, the focus has been on meeting the challenges
of quantum gravity proper that have been with us for decades: the
resolution of space-time singularities of classical general
relativity, extensions of early universe scenarios to the Planck
regime, introduction of $n$ point functions and development of a
scattering theory in a background independent context, \ldots .

In terms of motivation, quantum gravity programs have a strong
similarity with general relativity. Einstein's main goal was to
reconcile two fundamental and successful theories; Newtonian gravity
and special relativity. He was disturbed by the deep tension between
their underlying principles and convinced that the apparent conflict
arose because they were special cases of a deeper and grander
theory. He was not trying to modify Newtonian gravity to explain any
observational discrepancy such as the difference between the
calculated and observed values of the perihelion of mercury. Work in
non-perturbative quantum gravity is driven by the same spirit: The
goal is to find the grander, deeper theory from which general
relativity and quantum field theory arise as special cases. But
because there is neither observational data to guide us nor, alas!,
a second Einstein to leap over this profound limitation, we have not
seen definitive paradigm shifts. We do see solutions to some of the
important problems but they have come in pieces; a compelling global
picture has not emerged. Perhaps the lasting legacy of efforts to
date will be that they have provided a host of novel and powerful
mathematical tools, and sufficiently sharpened the conceptual
issues, to bring out deep tensions between gravity and the quantum
that we were blissfully unaware of. The new and incisive questions
that have arisen are likely to be our best guides in the coming
years.

\section{Epilogue}
\label{s6}

I have presented only an illustrative sample of advances our field
has made since GR3. Clearly, there is a lot to be proud of. As we
stand at this threshold of the Centennial of general relativity, is
there a simple phrase that succinctly encapsulates our collective
sentiment? I think there is. Not surprisingly, it comes from
Einstein himself.

Einstein presented his calculation of the perihelion advance of
mercury to the Prussian Academy on November 18th, 1915. Just ten
days later, he wrote to Arnold Sommerfeld in Munich saying: ``During
the last month, I experienced one of the most exciting and most
exacting times of my life and true enough also one of the most
successful ...". He then went on to explain `` Now the marvelous
thing which I experienced was the fact that not only did Newton's
theory result as first approximation but also the perihelion of
mercury ($43''$ per century) as second approximation ...".
Apparently, Sommerfeld was puzzled by this uncharacteristic
enthusiasm of Einstein's. So, on February 8th, 1916 Einstein wrote
back saying
\begin{quote}
\emph{``Of general theory of relativity, you will be convinced once
you have studied it. Therefore I am not going to defend it with a
single word."}
\end{quote}
A century has passed and yet this assessment continues to capture
our core reaction to general relativity.\\

But it is equally interesting that fundamental issues still remain
\emph{even in the classical theory}. For example, we still do not
have a satisfactory notion of a \emph{dynamical black-hole}, one
that is not teleological, one that can be used to say with
confidence that there is no black hole in the room you are now
sitting in. The notion of an event horizon allows this possibility
because event horizons can form and grow even in regions where the
space-time metric is flat!%
\footnote{For further elucidation of limitations of this notion,
see. e.g., \cite{akrev}. Finding a satisfactory characterization of
dynamical black holes continues to be an active topic of research.
For example, at GR20 we heard of the interesting idea of `the core
of a black hole'.}
Similarly, there is a fundamental conceptual question about
gravitational waves. Observationally, the universe seems to have a
\emph{non-zero}, positive cosmological constant, $\Lambda$. But we
do not have a satisfactory theory of gravitational radiation if
$\Lambda >0$. In particular, we do not know the analog of the `Bondi
news'. Even today, 50 years after GR3, there is no reliable
framework for us to echo Bondi and say with confidence: \emph{Yes,
there is a gauge invariant characterization of gravitational waves
in full general relativity with a positive cosmological constant,
and they carry positive energy.} We do not even know a precise
boundary condition that would correctly capture the idea that there
is no incoming radiation in space-times describing isolated systems.
Nor do we have the analog of the beautiful positive energy theorems
if $\Lambda >0$.

And of course \emph{many} fundamental issues at the interface of
general relativity and quantum physics continue to be hotly debated:
the issue of information loss, the fate of classical singularities
in the quantum theory, physics of the very early universe in the
Planck epoch, the initial conditions and measurement theory in
cosmology, \ldots .

These fundamental issues and deep tensions offer great opportunities
for the future. So do the challenges of testing general relativity
in the truly strong field regime and the tremendous potential of
gravitational wave astronomy. Thanks to all these opportunities, the
field of gravitational science is becoming ever more fertile. As we
approach the centennial of general relativity, a transformation has
already been set in motion, one that will take us well beyond
Einstein's vision. Research in the type of analytical general
relativity that dominated GR3 has been receding. The field has moved
into new areas: geometric analysis, cosmology, relativistic
astrophysics, computational science, high energy physics,
gravitational wave astronomy and particle astrophysics. In future,
the emphasis will be on using relativistic gravity to provide us a
more holistic view of the cosmos (see e.g. \cite{amon}). As this
transformation unfolds, there will be numerous fresh insights,
unforeseen advances, new puzzles and even paradigm shifts. In
another half a century, the GR37 conference will be held (perhaps
again in Warsaw!). It will surely be at least as engaging and
stimulating a conference as GR3 and GR20 have been. But it will be a
\emph{very} different one!

\section*{Acknowledgments}

I would like to thank Marek Demianski for sending me link to the
on-line version of the GR3 conference proceedings, Charles Lawrence
for detailed discussions on the Planck mission and John Friedman for
correspondence related to the development of relativistic
astrophysics. This work was supported by the NSF grant PHY-1205388
and the Eberly research funds of Penn state.


\begin{thebibliography}{99}

\bibitem{gr3} \emph{Proceedings `on theory of gravitation'
    conference in Warszawa and Jablona}, edited by L. Infeld
    (Guthier-Villars, Paris, and PWN E\'ditions Scientifiques de
    Pologne, Warszawa (1964)).
\bibitem{er} D. Kenneflick, \emph{Who is afraid of the referee?
    Einstein and gravitational waves, Talk at Stony Brook}, October 2005,
    http://dafix.uark.edu/~danielk/Physics/gravwave.html
\bibitem{planck} Planck collaboration, Planck 2013 results. I.
    Overview of products and scientific results,
    \texttt{arXiv:1303:5062 v1}.
\bibitem{jm}J.~Mitchell, \emph{On the Means of Discovering the
    Distance, Magnitude, \&c. of the Fixed Stars, in Consequence of the
    Diminution of the Velocity of Their Light, in Case Such a
    Diminution Should be Found to Take Place in any of Them, and
    Such Other Data Should be Procured from Observations, as Would
    be Farther Necessary for That Purpose}, Phil. Trans. Royal
    Soc. (London), Vol. 74, 35-57, (1783).
\bibitem{akrev}A. Ashtekar and B.~Krishnan, Isolated and dynamical
    horizons and their applications, Liv. Rev. Rel.7:10 (2004).
\bibitem{amon} http://amon.gravity.psu.edu/index.shtml



\end{thebibliography}
\end{document}